\documentclass[aps,twocolumn,pre,showpacs,nofootinbib]{revtex4}
\usepackage{graphicx}

\newcommand{\MCtwo}{Microtechnology and Nanoscience, MC2, 
Chalmers University of Technology, SE-412 96 G{\"o}teborg, Sweden}
\newcommand{\Hule}{Huleb\"acksgymnasiet, 
Idrottsv\"agen 2, SE-435 80 M\"olnlycke, Sweden}

\newcommand{\gasgrap}{{\mbox{\scriptsize gas, fix-cell}}}
\newcommand{\adsgrap}{{\mbox{\scriptsize ads, fix-cell}}}

\newcommand{\LDA}{{\mbox{\scriptsize LDA}}}

\newcommand{\tot}{{\mbox{\scriptsize tot}}}
\newcommand{\nl}{{\mbox{\scriptsize nl}}}

\newcommand{\CH}{{\mbox{\scriptsize C-H}}}
\newcommand{\CCl}{{\mbox{\scriptsize C-Cl}}}
\newcommand{\CBr}{{\mbox{\scriptsize C-Br}}}

\begin{document}

\title{A van der Waals density functional study of chloroform and 
bromoform on graphene}

\author{Joel {\AA}kesson}\affiliation{\Hule}
\author{Oskar Sundborg}\affiliation{\Hule}
\author{Olof Wahlstr\"om}\affiliation{\Hule}
\author{Elsebeth Schr{\"o}der}\thanks{Corresponding author}%
\email{schroder@chalmers.se}%
\affiliation{\MCtwo}

\date{May 28, 2012}

%%%%%%%%%%%%%%%%%%%%%%%%%%%%%%%%%%%%%%%%%%%%%%%%%%%%%%%%%%%%%%%%%%%%%%%%%%%%%%
\begin{abstract}
A calculational study of the trihalomethanes chloroform (CHCl$_3$) 
and bromoform (CHBr$_3$) adsorbed on graphene is presented. 
The study uses the van der Waals density functional method vdW-DF to obtain 
adsorption energies and adsorption structures for these molecules of
environmental concern. 
In this study chloroform is found to adsorb with the H atom pointing away from
graphene, with adsorption energy 357 meV (34.4 kJ/mol). For bromoform the
calculated adsorption energy is 404 meV (39.0 kJ/mol).
The corrugation of graphene as seen by chloroform is small, the difference
in adsorption energy along the graphene plane is less than 6 meV. 
\end{abstract}

\maketitle

%%%%%%%%%%%%%%%%%%%%%%%%%%%%%%%%%%%%%%%%%%%%%%%%%%%%%%%%%%%%%%%%%%%%%%%%%%%%%%

\section{Introduction} 

Trihalomethane (THM) molecules are small molecules that are similar to 
methane (CH$_4$) but with three of the H atoms replaced by halogens 
(F, Cl, Br, I, At). The most welknown THM is trichloromethane (CHCl$_3$), 
also known as chloroform. THMs are of environmental concern as they are 
toxic to human health \cite{tox,tox2}.  
The human body adsorbs THMs by inhalation and by passage through the skin,
but the main contribution to human exposure arises from the consumption of 
chlorinated drinking water \cite{2006-report}.

Of the THMs, chloroform is found in the highest concentration in the
environment. Chlorine  used for water disinfection reacts with organic 
material in the water, forming a number of THMs as byproducts:
mainly chloroform, but also THMs with one or more Br atoms.
The THMs toxity motivates a search for an effective process for selective
extraction.
Carbon materials, such as activated carbon or carbon nanotubes are used 
or have been suggested for the use in adsorbing filters for removing 
THMs from the drinking water after the disinfection, but before 
intake \cite{2006-report,girao}.

We here study how a chloroform molecule adsorbs on the simplest of 
carbon materials, graphene. By use of density functional theory (DFT)
calculations we determine the energy gained at adsorption and compare with
the adsorption energies of similar molecules, like methane and 
tribromomethane (CHBr$_3$), also called bromoform.
For these calculations we apply the first-principles van der Waals (vdW) 
density-functional method vdW-DF \cite{Dion,Thonhauser}. 

Chloroform on carbon materials has previously been 
studied in a few studies by use of theory. DFT has been used for chloroform 
on benzene in a study employing the vdW-DF method \cite{hooper} 
and for a study of chloroform on carbon nanotubes with use of the local 
density approximation (LDA) \cite{girao}. For experiments, there is a 
century long tradition of studies of chloroform because it was frequently 
used as a solvent and as an anasthetic. However, adsorption studies 
on carbon materials, provinding adsorption (or desorption) energies, are 
more recent \cite{hertel,girao,rybolt}.

The outline of the rest of the paper is as follows: In Section II we 
describe the computational method and our system of chloroform and 
graphene. In Section III we present our results and discussions, 
and Section IV contains our summary. 
%%%%%%%%%%%%%%%%%%%%%%%%%%%%%%%%%%%%%%%%%%%%%%%%%%%%%%%%%%%%%%%%%%%%%%%%%%%%%%

\section{Method of computation}
THMs are molecules with a central C atom and four other atoms surrounding 
the C atom approximately evenly distributed. Of these, one atom is an 
H atom and the three other atoms are halogens. In this paper we analyze 
the adsorption on graphene of chloroform and bromoform, 
for which the three halogen atoms are Cl or Br atoms, respectively.

We use DFT with the vdW-DF method \cite{Dion,Thonhauser} to determine 
the adsorption energy and atomic structure. Our calculations are carried 
out fully self-consistently \cite{Thonhauser}. We use the DFT code 
GPAW \cite{gpaw} with vdW-DF \cite{Dion,Thonhauser}
in a Fast-Fourier-Transform implementation \cite{soler}.
The GPAW code is an all-electron DFT code based on
projector augmented waves \cite{Blochl} (PAW).
 
\begin{figure}[bt]
\begin{center}
\includegraphics[width=0.3\textwidth]{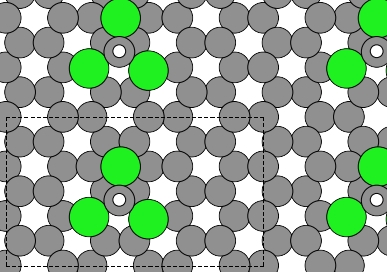}\\[0.7em]
\includegraphics[width=0.3\textwidth]{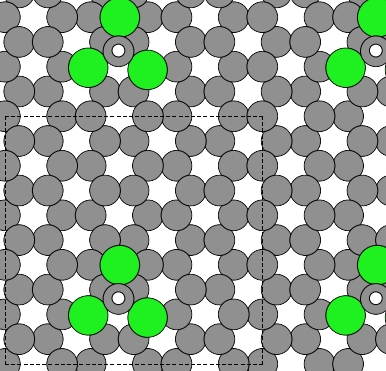}\\
\includegraphics[width=0.4\textwidth]{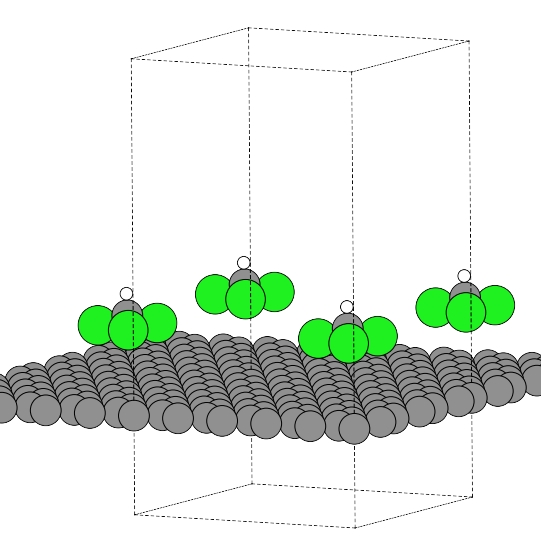}
\caption{Illustration of chloroform adsorbed on graphene for 
the $3\sqrt{3}\,a_g\times 3\,a_g$ unit cell (top panel) and
the $3\sqrt{3}\,a_g\times 5\,a_g$ unit cell (middle and bottom panels).
The unit cell is outlined by the broken lines. Also shown are atoms of
part of the 
neighboring repeated unit cells, illustrating the separation of the
repeated images of the chloroform molecule. 
C atoms are represented by medium size gray circles, H atoms by
small white circles, and Cl atoms by large green circles.
}
\label{fig:chloroform}
\end{center}
\end{figure}

Figure \ref{fig:chloroform} illustrates the adsorbed chloroform molecule
on graphene, and the periodically repeated orthorhombic unit cells used 
in our calculations. 
We show the adsorption configuration that has the H atom pointing away 
from graphene. A previous vdW-DF study \cite{hooper} of chloroform with 
benzene discussed the C-H/$\pi$ interaction in a dimer with the H atom 
of chloroform pointing towards the center of the aromatic benzene ring.
In order to connect to that study we also carry out calculations with 
the H atom pointing towards graphene and find that configuration to 
have a smaller (less favorable) adsorption energy than configurations 
with the H atom
pointing away from graphene. In this work we therefore focus on 
adsorption configurations with Cl (or Br) atoms closest to graphene 
as shown in Figure \ref{fig:chloroform}.

We define the adsorption energy $E_a$ as
the difference in total energy when the molecule is adsorbed on
graphene ($E^\tot_\adsgrap$) and when it is in the gas phase
far away from graphene ($E^\tot_\gasgrap$)
\begin{equation}
-E_a=E^\tot_\adsgrap-E^\tot_\gasgrap.
\label{eq:Ea}
\end{equation}
Here we follow the sign convention that yields positive values of $E_a$ 
for systems that bind.
The two total-energy terms in (\ref{eq:Ea}) are both calculated with the
adsorbant and graphene within one unit cell, the unit cell having the 
same size in both calculations.%
\footnote{In previous work the change of the adsorbant structure from 
the deformed structure (after removal from graphene) into
the gas phase structure was sometimes \protect\cite{nalkanes} 
calculated with a GGA approximation of $E_{xc}$
because that approximation is less sensitive to
changes in grid positions \protect\cite{nalkanes,PAHgg,PAHgraphite}.
However, with the fine real-space grid used here
and the rather small adsorbant molecules
the total energy difference between the deformed and the gas phase structure
of chloroform or bromoform is less than 1 meV. We therefore here solely use
vdW-DF in the calculations.}
Because $E^\tot_\gasgrap$ is calculated with the same adsorbant-adsorbant 
separation as $E^\tot_\adsgrap$ the direct lateral interaction
is subtracted from our results \cite{adenine,slidingrings}.

The optimal positions of the atoms within chloroform (and bromoform) 
are determined by minimization of the Hellmann-Feynman forces
acting on the chloroform (or bromoform) atoms, when adsorbed on to 
graphene (``\adsgrap") and when away from graphene (``\gasgrap").
We use the molecular-dynamics optimization method ``fast inertial
relaxation engine" (FIRE) \cite{fire} with the requirement that the 
size of the remaining force on each atom is less than 0.01 eV/{\AA}.
The positions of the graphene atoms are left unchanged.
The Hellmann-Feynman forces are derived from gradients in the electron 
density $n$. The optimization yields the bond lengths and angles in the
molecules (after adsorption and in the gas phase) and the optimal 
position of the molecule with respect to graphene.
The potential well for the
molecule near graphene is very shallow. We therefore start the 
optimization process in several different lateral and vertical 
positions of chloroform. 
 
We use an orthorhombic unit cell of size 
$3\sqrt{3}\,a_g\times 3\,a_g \times 23.0$ {\AA} 
for most of our calculations, but we also test the adsorption energy
convergence with unit cell size, using a unit cell of size
 $3\sqrt{3}\,a_g\times 5\,a_g \times 23.0$ {\AA}. Here 
$a_g= \sqrt{3}\, a_0$, and $a_0=1.43$ {\AA} is the clean graphene lattice
constant as found earlier by relaxing the lateral 
size of the unit cell \cite{nalkanes}. 
In the direction perpendicular to graphene there is 
$\sim 12$ {\AA} of vacuum above
chloroform when desorbed (and $\sim 19$ {\AA} when adsorbed)
to avoid
vertical interaction across unit cell boundaries.

The (valence) electron density and wave functions are represented on
evenly distributed grids in space.
To ensure a good accuracy in our calculations we choose the density grid
to have approximately 0.12 {\AA} grid point separation in all three
directions \cite{Ziambaras}.
In all calculations we use a $2\times$2$\times$1 Monkhorst-Pack
$k$-point sampling of the Brillouin zone.

In DFT the total energy $E^\tot[n]$ is given as a functional of the
electron density $n$. DFT is in principle exact, but in practise 
the exchange-correlation part $E_{xc}$ of the total energy must
be approximated.
The exchange-correlation energy may also be written as a sum of
exchange and correlation contributions, $E_{xc}=E_x+E_c$. 

In the vdW-DF method the $E_x$ is chosen as the exchange part of a 
generalized gradient approximation (GGA). In the 
original version of vdW-DF \cite{Dion} (which we use here)  
the exchange chosen is that of the revPBE approximation \cite{revPBE}.
The correlation energy $E_c$ 
is split \cite{IJQC} into a nearly-local part $E_c^0$ and a part that
includes the most nonlocal interactions $E_c^\nl$, 
\begin{equation}\label{eq:1}
E_c[n] = E_c^0[n]+E_c^\nl[n] \,.
\end{equation} 
In a homogeneous system the term $E^0_c$ is the correlation $E^\LDA_c$ 
obtained from the local density approximation (LDA),
and in general \cite{Dion} we approximate $E^0_c$ by $E^\LDA_c$.
The term 
\begin{equation}\label{eq:2}
E_c^\nl[n] =
\frac{1}{2}\int\int d\mathbf{r}\,d\mathbf{r}'\,n(\mathbf{r})\phi(\mathbf{r},\mathbf{r}')n(\mathbf{r}')
\end{equation}
describes the dispersion interaction and vanishes for a homogeneous system.
It is
given by a kernel $\phi$ which is explicitly stated in Ref.~\onlinecite{Dion}.

The term $E_c^\nl$ is sensitive
to changes in the local real space grid, for example
when the nuclei positions are translated by a distance
not corresponding to an integer
number of real-space grid points \cite{PAHgg,Ziambaras,NTgg}.
For calculations involving small (few meV) energy differences,
like the calculations for the potential energy  surface introduced
in Section \ref{sec:III.D} we therefore keep positional changes
of the rigid molecules to an integer number of grid points.

Finally, we
make sure that each of the total-energy calculations
are accurately converged to a change in the total energy of 
less than 0.1 meV per unit cell, or less than approximately 10$^{-6}$ eV
per atom in the unit cell, in the last three DFT 
iterations.
This choice of allowed change is significantly smaller than the default
of GPAW and is essential in order to discuss energy changes in this
adsorption system where relevant differences in total energies are 
down to the meV scale.

\begin{table*}
\caption{\label{tab:bonds} 
Bond lengths and bond angles of the gas phase and the adsorbed phase 
chloroform and bromoform molecules. Adsorption structures are for the
``H-up" structure (see text).
For the theory results the vdW-DF method is used \protect\cite{Dion,Thonhauser}.
Error bars on results from the NIST data base \protect\cite{nist} are 
not available for chloroform.
}

\begin{tabular}{lrrrrr}
 \hline\hline
                       &\multicolumn{1}{c}{NIST}& \multicolumn{3}{c}{This work}&    \\
                       &        & \multicolumn{3}{c}{\line(1,0){150}} &   \\
     &\multicolumn{1}{c}{gas phase}& gas phase &ads.\ phase& $|\Delta(\mbox{gas}-\mbox{ads.})|$&$|\Delta(\mbox{NIST}-\mbox{gas})|$ \\
  \hline
  chloroform           &  &  &  &  &  \\
  $d_\CH$     ({\AA})  & 1.073           & 1.087 & 1.087 & $<0.001$ & 0.01 (1\%) \\
  $d_\CCl$    ({\AA})  & 1.762           & 1.798 & 1.801 & 0.003    & 0.03 (2\%) \\
  $\angle$Cl-C-Cl (deg)& 110.92          & 111.4 & 111.1 & 0.3      & 0.5 ($<1$\%) \\
  $\angle$H-C-Cl (deg) & 107.98          & 107.7 & 107.7 & $<0.1$   & 0.3 ($<1$\%) \\
  bromoform            &  &  &  &  &  \\
  $d_\CH$     ({\AA})  & $1.11\pm0.05$   & 1.085 & 1.085 & $<0.001$ & 0.02 (2\%) \\
  $d_\CBr$    ({\AA})  & $1.924\pm0.005$ & 1.974 & 1.977 & 0.003    & 0.05 (3\%) \\
  $\angle$Br-C-Br (deg)& $111.7\pm0.4$   & 111.5 & 111.8 & 0.3      & 0.2 ($<1$\%) \\
  $\angle$H-C-Br (deg) & $107.2\pm0.4$   & 106.9 & 106.9 & $<0.1$   & 0.3 ($<1$\%) \\
\hline
    \end{tabular}
\end{table*}

\begin{table*}
\caption{\label{tab:energies} Adsorption energies $E_a$ from theory 
and experiment, distance of molecular C atom from the plane of graphene, 
$d_R$, adsorption configuration (H atom sticking up or down), and unit cell used in calculations.
The method vdW-DF of Refs.\ \protect\cite{Dion,Thonhauser} is used when not
noted otherwise.
The adsorption structures for the experiment results are not known to us
and for each molecule experiments are entered in the first entry line. 
We use orthogonal unit cells and a graphite lattice vector $a_g=\sqrt{3} \,a_0$ 
with $a_0=1.43$ {\AA}.
}

\begin{tabular}{llccccc}
 \hline\hline
                       &        & \multicolumn{4}{c}{This work}           &\multicolumn{1}{c}{Experiments}\\
                       &        & \multicolumn{4}{c}{\line(1,0){150}}     &\\
                       &Structure& Unitcell           &\multicolumn{2}{c}{$E_a$}&$d_R$&$E_a$ \\
                       &        &                     &[kJ/mol]&[meV]&[{\AA}] & [kJ/mol]\\
  \hline
  chloroform           & H up   & $3\sqrt{3}\times 3$ & 34.4 & 357 & 4.20    & 54$\pm 3^a$, 36.4$^b$   \\
                       & H up   & $3\sqrt{3}\times 5$ & 34.3 & 356 & 4.20    &                   \\
                       & H down & $3\sqrt{3}\times 3$ & 33.5 & 347 & 3.48    &                   \\
  bromoform            & H up   & $3\sqrt{3}\times 3$ & 39.0 & 404 & 4.32    &           \\
  methane$^c$          &        & $3\sqrt{3}\times 3$ & 14.6 & 152 & 3.64    &13.6$^d$, 17$\pm 1^a$   \\
\hline
\multicolumn{7}{l}{${}^a$Thermal desorption spectroscopy measurements (at one monolayer), Ref.~\protect\cite{hertel}.}\\
\multicolumn{7}{l}{${}^b$Single atom gas chromotography, Ref.~\protect\cite{rybolt}.}\\
\multicolumn{7}{l}{${}^c$Theory results from previous vdW-DF study with the same settings, Ref.~\protect\cite{nalkanes}.}\\
\multicolumn{7}{l}{${}^d$Temperature programed desorption (results extrapolated to isolated adsorbant), Ref.~\protect\cite{tait2006}.}

    \end{tabular}
\end{table*}

%%%%%%%%%%%%%%%%%%%%%%%%%%%%%%%%%%%%%%%%%%%%%%%%%%%%%%%%%%%%%%%%%%%%%%%%%%%%%%% 
\section{Results and discussion}

Besides the total energies, in our calculations we also determine the 
molecular structure of the chloroform and bromoform molecules, both in 
the dilute gas phase and in the adsorbed phase. We compare the bond 
lengths and angles of the gas phase molecules with experimental values, 
and we determine the changes that occur when the molecules are adsorbed.

The chloroform and bromoform molecules may adsorb in various orientations 
and positions on graphene. We determine the optimal orientation and 
adsorption distance, and discuss the influence of the adsorption 
position on the adsorption energy. Further, we test the convergence of 
the adsorption energy (\ref{eq:Ea}) with respect to lateral size of 
unit cell, i.e., 
the length of the smallest molecule-to-molecule distance.  

\subsection{Structure of the desorbed and adsorbed molecules}
For the desorbed molecules the atomic positions within the molecule are
determined in the process of determining the total energy of the 
system of a molecule far away from graphene. 
The total energy of these gas phase structures, along with the
clean graphene total energy, are used as the desorbed system energy
with which the adsorbed system total energies are compared. The 
sum of total energies of the desorbed system is used as the zero 
point of the adsorption energy. In practise
for our desorption calculations we keep graphene and chloroform
(or bromoform) within the same unit cell, but far apart.

{}From the atomic positions in the gas phase the bond lengths and bond angles 
within the
chloroform and bromoform molecules are extracted and listed in 
Table \ref{tab:bonds}.
We find that the C-Cl bonds are slightly shorter than the C-Br bonds, 
a result that is  
expected because Br is a larger halogen atom than Cl. The bond values we
find are in reasonable agreement with the experimental values listed in
the NIST database \cite{nist}.
All bond lengths are within 1-3\% of the experimental values.
Like for the experiments, the bond angles 
$\angle$Cl-C-Cl (and $\angle$Br-C-Br) are a few degrees larger than 
the $\angle$H-C-Cl (and $\angle$H-C-Br) angles.
For the bond angles our results deviate less than 1\% from experiment
(Table \ref{tab:bonds}).

When adsorbed, the relative positions of the atoms in the adsorbant change.
However, as we have also found in other small physisorbed 
molecules \cite{nalkanes,methanol}, the changes are very small. 
Table \ref{tab:bonds} lists the changes.

\subsection{Adsorption energies}
Our main calculations are optimizations of the adsorbed
chloroform molecule when it is positioned with the H atom
pointing away from graphene (``H up"), as illustrated in 
Figure \ref{fig:chloroform}. 
We find adsorption energies in the 350--360 meV range
depending on the precise position on graphene. 
For the position in Figure \ref{fig:chloroform}
we find $E_a=357$ meV. As discussed below, the 
differences in adsorption energies in the various 
positions on graphene are small.

Like the structural changes in the chloroform and bromoform
molecule upon adsorption, the energetic changes are 
small. In fact, we find the contribution to the adsorption 
energy from the deformation of the adsorbant to be about 0.2 meV, less 
than the accuracy of our results ($\sim 1$ meV). 

Table \ref{tab:energies} also lists the adsorption energies obtained
through experimental measurements. These find that chloroform 
binds stronger to graphene than our results, with a deviation
of our results from the experiments 36\% for thermal desorption
spectroscopy \cite{hertel} or 5\% for single atom chromotography
\cite{rybolt}.

For our calculations we also report in Table \ref{tab:energies}
the distance $d_R$ between the C atom of chloroform (or bromoform)
and graphene, in the adsorbed configuration.

The majority of our calculations are carried out in a unit cell of 
lateral size $3\sqrt{3} \,a_g \times 3\, a_g$, leading to a smallest
(center-of-mass) molecule-to-molecule distance 7.43 {\AA} for 
the periodically repeated images, or the distance 5.00 {\AA} between
two closest Cl atoms each residing on a different chloroform molecule. 
In our calculational procedure described by (\ref{eq:Ea})
we subtract any direct interactions between the molecules because
both terms are calculated with the same lateral distance to neighboring
molecules (i.e., with the same lateral size of the unit cell). Indirect
interactions leading to changes in adsorption energy, could for example
arise via a possible small deformation of the electron distribution
on graphene, or via a possible but tiny deformation of atomic
structure on chloroform. In order to check such possible effect
we also calculated the adsorption of chloroform in a unit cell
with 12.38 {\AA} molecule-to-molecule distance, the 
$3\sqrt{3} \, a_g \times 5\, a_g$ unit cell. 
As evident from the results in Table \ref{tab:energies} there is
very little difference between results of the small and large 
unit cell, confirming that the $3\sqrt{3}\, a_g \times 3\, a_g$
unit cell size is sufficient for size-convergence,
provided that direct molecule-molecule interactions 
across unit cell boundaries are cancelled
like we do here.

In order to check whether the chloroform configurations with the 
H atom pointing to graphene are energetically more favorable than the 
``H up" configurations we also carry out a number of ``H down" calculations.
We calculate $E_a$ for the configuration with the H atom above 
the center of an aromatic ring in graphene and in a number of
configurations with the H atom near or above a graphene C-C bridge.
We find the configuration with H above the center of an aromatic ring to be 
the most favorable of the H-down configurations (H centered configuration
listed in Table \ref{tab:energies}) by up to 12 meV, but less favorable 
that configurations with the H atom pointing away from graphene.

Previously, the dimer of chloroform with benzene has been studied using 
vdW-DF \cite{hooper}. In that study the interaction between an
aromatic $\pi$-system (represented by benzene) and an aliphatic C-H
group (in chloroform) was at focus, and accordingly the orientation of the 
chloroform was chosen to have the H-atom pointing towards the (center of)
the benzene molecule. In Ref.\ \cite{hooper} the binding energy 5.11 kcal/mol
(21.4 kJ/mol or 222 meV/dimer) was found, with the distance from chloroform-C to 
the benzene-plane $d_R= 3.6$ {\AA}.  
The vdW-DF results of that study were compared with coupled cluster 
[CCSD(T)] calculations \cite{ringer} with binding energy 5.60 kcal/mol 
(23.4 kJ/mol or 243 meV/dimer) at $d_R= 3.2$ {\AA}.
In our calculations the interaction of chloroform is with the full
graphene plane (as far as the vdW forces reach) and not only with
the benzene molecule, we should therefore expect a larger 
interaction energy than for the molecular dimer. Indeed, with an
adsorption energy of 347 meV (Table \ref{tab:energies})
we do find stronger binding than in the dimer case, stronger by
about 123 meV, and with an adsorption position closer to the aromatic 
(graphene or benzene) ring
by about 0.1 {\AA} compared to the vdW-DF calculation and 
further away by approximately 0.3 {\AA} compared to the CCSD(T) calculation. 

In Ref.\ \cite{girao} the adsorption of chloroform on to the (5,0) and (8,8)
single walled carbon nanotubes (SWCNT) was addressed with the use of the LDA 
approximation to DFT.
Adsorption energies 200 meV [for the (5,0) SWCNT] and 150 meV [for the (8,8) SWCNT]
were found. However, even though LDA seems to bind vdW materials  
it cannot be used for the inclusion
of vdW interactions, as already pointed out by Harris \cite{harris} and discussed 
also in Refs.\ \cite{adenine,murray}. The LDA results of Ref.\ \cite{girao} are
therefore not further discussed here.  

\subsection{Adsorption of bromoform} 
The Br atom is similar to Cl but is heavier, i.e., has more electrons 
and larger molecular polarizability. We therefore
expect a stronger binding to graphene than for
chloroform.
Indeed, our results in Table \ref{tab:energies}
show the adsorption energy to be 47 meV larger than that for chloroform. 
The adsorption distance
as measured from the bromoform C atom is larger, reflecting the fact that
the Br atom has a larger volume.
The same trend, in the opposite direction, is seen for methane \cite{nalkanes}:
the H atom has less electrons than Cl, the binding is less strong 
(by $\sim$200 meV compared to chloroform) and the binding distance 
$d_R$ is smaller because H has less volume than Cl.

For the bromoform result shown in
Table \ref{tab:energies} we used the same position on graphene as
illustrated in the top panel of
Figure \ref{fig:chloroform} as the starting point for the
structural relaxations.

\begin{figure}
\begin{center}
\begin{tabular}{cc}
$\vcenter{\hbox{\includegraphics[width=0.38\textwidth]{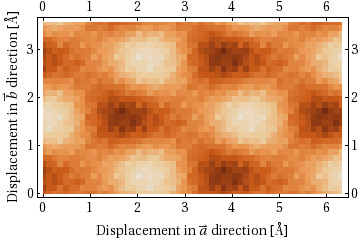}}}$
& 
$\vcenter{\hbox{\includegraphics[width=0.19\textwidth, angle=90]{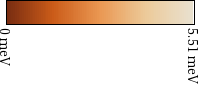}}}$
\end{tabular}

\caption{\label{fig:PES}Potential energy surface (PES) for chloroform on graphene in
the ``H-up'' adsorption structure.  Chloroform
is here kept at a distance $d=4.20$ {\AA} from graphene, 
measured from the chloroform C atom to the plane of graphene.
The PES plot scans 1/4 of the $3\protect\sqrt{3}\, a_g \protect\times 3\, a_g$
unit cell, with origo taken as the chloroform position of
the top panel in Figure
\protect\ref{fig:chloroform} 
and with the same orientation as that panel.
The energy scale measures the deviation from the global minimum.
}
\end{center}
\end{figure}

\subsection{Chloroform potential energy surface\label{sec:III.D}}
In Figure \ref{fig:PES} we show the potential energy surface (PES) 
for chloroform. The figure is obtained from translating chloroform
along graphene with the chloroform 
C atom kept at a fixed distance $d=4.20$ {\AA} from
the graphene plane. 
In Figure \ref{fig:PES} origo corresponds to the
position shown in the top panel of Figure \ref{fig:chloroform}.
We find that the 
variation in adsorption energy along graphene is small, with an
energy difference of less than 6 meV when $d=4.20$ {\AA}. 
This illustrates that
the corrugation of graphene, as experienced by the adsorbed
chloroform molecules, is very small, and it takes only very 
little kinetic energy to overcome the barriers for lateral
motion on graphene. Therefore, the concept of ``adsorption sites"
is not relevant in these physi\-sorp\-tion studies \cite{h2cu,h2cu2}.
In effect, the chloroform is free to 
move along graphene at all temperatures relevant in 
practical applications mentioned in the introduction.

{}From Figure \ref{fig:PES} we also find that the position used for
the calculated adsorption energy in Table \ref{tab:energies} is
close to but not quite the optimal position, 
albeit not the energetically worst position 
either. In any case, the effect of our choice of position is a few meV,
or less than 1\% of the adsorption energy. 

In all calculations for the PES the relative
positions of the atoms within the adsorbants are kept the same, thus
no further relaxation of the molecular structure is allowed. However,
since the change in molecular structure when desorbing is on the
sub-meV level, 
and less than the accuracy of our calculations, we expect the 
differences in molecular structure in the various adsorption positions 
to be even smaller and the effect on the adsorption energy to thus be 
negligible. 

All data points for the PES
are from adsorbant positions translated an integer number of grid points
along the surface, that is, on a uniform orthogonal grid with 0.12 {\AA}
between grid lines. We restrict the possible positions of chloroform
 in order to keep the atomic positions relative to the 
density grid the same in all calculations, thus avoiding any 
effects of the sensitivity of the vdW-DF results on the positioning
on the grid.
This is potentially important because the PES is mapping a very shallow
energy landscape. 

The data for the PES are calculated for all grid positions
within a $\sqrt{3}\, a_g \times 1\, a_g$ part of the 
$3\sqrt{3}\, a_g \times 3\, a_g$ 
unit cell, that is, by scanning over twice the area of the graphene 
primitive cell (which has only two C atoms).
Therefore, when translating the chloroform molecule over the 
$\sqrt{3}\, a_g \times 1\, a_g$ area we calculate two data points 
for each unique adsorption position.
To lower the sub-meV noise we use the average of the total energies in the
two equivalent positions for the plot in Figure \ref{fig:PES}. 
For clarity we also include in Figure \ref{fig:PES} repetitions of 
the calculations in the two lateral directions.
 
In Figure \ref{fig:PES} only results of the fixed distance $d=4.20$ {\AA}
are shown, but our calculations include also distances $4.20\pm 0.12$ {\AA}
in the same lateral positions as used in Figure \ref{fig:PES}. This 
corresponds to moving the molecule one full grid spacing closer to
or further away from graphene and redoing the PAS calculations. 
However, all adsorption energies at those 
distances are smaller than any of the adsorption energies at the 
$d=4.20$ {\AA} distance, and it is clear that the optimal distance 
$d_R$ from graphene varies much less 
than 0.12 {\AA} when moving along graphene.
Although the vertical part of the adsorption potential is shallow compared
to covalent and ionic binding the lateral part is even more shallow.

The differences in smallest and largest total energy within each of the three 
chloroform-C-to-graphene distances are 7.7 meV, 4.8 meV, and
2.9 meV for the distances $d=4.08$ {\AA}, 4.20 {\AA}, and 4.32 {\AA}, 
respectively. Thus the corrugation of graphene, as seen by chloroform,
becomes slightly more pronounced in positions closer to 
graphene, even though the corrugation is small
at all three distances considered here.

\subsection{Implications for environmental research.} 
Even water from untreated drinking water wells contains THMs
\cite{2006-report}. The THMs are spread in the environment since
chlorinated water is used for watering, and it leaks from swimming
pools and enters waste water. It is also produced by salt used on winter roads.
Chloroform is also relatively volatile and escapes from water into the
air with vapor (where inhalation may pose a health problem) or enters
through the skin, for example during showers or in swimming pools.
Absorption through the gastrointestinal tracts is fast and extensive,
with the majority of ingested chloroform recovered in expired air within
a few hours \cite{tox2}. With chloroform found in 11.4\% of public wells
in the U.S.\ \cite{2006-report} and chlorination still an important
candidate for improving the quality of drinking water in developing 
countries, the occurrence of THMs is a potential human health concern,
and methods to remove THMs after chlorination and before use of the water
should be improved.

Chlorination of water gives rise to a number of THMs as byproducts, 
mainly Cl- and Br-based THMs. In order for these to be removed from 
water by adsorption on to graphene it is necessary that the adsorption 
energy at least exceeds that of water on graphene, and that the adsorption 
energy is much higher than the barrier for thermal desorption at relevant 
temperatures, here around 300 K which corresponds to $\sim$26 meV. 
In a CCSD(T) study \cite{voloshina} the adsorption energy of water on 
graphene was found to be 135 meV, and in a recent vdW-DF study 
\cite{li2012} (utilizing a different exchange functional than here, the
optB86b exchange \cite{klimes2011}) the water adsorption energy
was found to be 140 meV. Both results are clearly smaller than 
our chloroform and bromoform adsorption energies that are in the range 
350--410 meV.
Thus, while not directly giving proof that graphene can be used for 
water filters after chlorination our results do suggest that this may 
be possible. 
 
%%%%%%%%%%%%%%%%%%%%%%%%%%%%%%%%%%%%%%%%%%%%%%%%%%%%%%%%%%%%%%%%%%%%%%%%%%%%%%
\section{Summary}
We present a study of the adsorption on graphene of the two THMs chloroform and
bromoform using the van der Waals density functional method vdW-DF. We find 
that chloroform and bromoform physisorb with their H atom pointing away from 
graphene, yielding adsorption energies 357 meV (34.4 kJ/mol) for chloroform
and 404 meV (39.0 kJ/mol) for bromoform. This suggests that these THMs bind
sufficiently strongly to graphene for graphene to be used in filtering of
chlorinated water to remove the THM byproducts. 

\acknowledgments
Partial support from the Swedish Research Council (VR) is gratefully acknowledged.
The computations were performed on resources provided by the Swedish 
National Infrastructure for Computing (SNIC) at C3SE.  
J{\AA}, OS, and OW carried out their part of the work
as a research course that is part of their high school education; 
the course was a collaboration between Huleb\"acksgymnasiet, 
G\"oteborg University, and Chalmers University of Technology, 
with project advisor ES and course supervisors Drs. Linda Gunnarsson and P\"ar Lydmark.

%%%%%%%%%%%%%%%%%%%%%%%%%%%%%%%%%%%%%%%%%%%%%%%%%%%%%%%%%%%%%%%%%%%%%%%%%%%%%%%%%%%


\begin{thebibliography}{99}

\bibitem{tox}
K. Foxall,
\textit{Chloroform Toxicological Overview,}
Health Protection Agency (U.K.) (2007).

\bibitem{tox2}
J.T. Du,
\textit{Toxicological Review of Chloroform,}
CAS No. 67-66-3,
U.S. Environmental Protection Agency Washington, DC (2001).

\bibitem{2006-report}
T. Ivahnenko and J.S. Zogorski,
\textit{Sources and Occurrence of Chloroform and Other Trihalomethanes
in Drinking-Water Supply Wells in the United States, 1986-2001},
Scientific Investigations Report 2006-5015,
U.S. Department of the Interior and U.S. Geological Survey (2006).

\bibitem{girao}
E.C. Gir\~{a}o, Y. Liebold-Ribeiro, J.A. Batista, E.B. Barros,
S.B. Fagan, J.M. Filho, M.S. Dresselhaus, and A.G.S. Filho,
Phys. Chem. Chem. Phys. \textbf{12}, 1518 (2010).

\bibitem{Dion}
M. Dion, H. Rydberg, E. Schr\"oder, D.C. Langreth, and B.I. Lundqvist,
Phys. Rev. Lett. \textbf{92}, 246401 (2004); \textbf{95}, 109902(E) (2005).

\bibitem{Thonhauser}
T. Thonhauser, V.R. Cooper, S. Li, A. Puzder, P. Hyldgaard, and D.C. Langreth,
Phys. Rev. B \textbf{76}, 125112 (2007).

\bibitem{hooper}
J. Hooper, V.R. Cooper, T. Thonhauser, N.A. Romero, F. Zerilli,
and D.C. Langreth,
ChemPhysChem \textbf{9}, 891 (2008).

\bibitem{hertel}
H. Ulbricht, R. Zacharia, N. Cindir, and T. Hertel,
Carbon \textbf{44}, 2931 (2006).

\bibitem{rybolt}
T.R. Rybolt, D.L. Logan, M.W. Milburn, H.E. Thomas, and A.B. Waters,
J. Colloid Science \textbf{220}, 148 (1999).

\bibitem{gpaw}
Open-source, grid-based PAW-method DFT code \textsc{gpaw},
\texttt{http://wiki.fysik.dtu.dk/gpaw/} ;
J.J. Mortensen, L.B. Hansen, and K.W. Jacobsen,
Phys. Rev. B \textbf{71}, 035109 (2005).

\bibitem{soler}
G. Rom\'an-P\'erez and J.M. Soler, Phys. Rev. Lett. \textbf{103}, 096102 (2009).

\bibitem{Blochl}
P.E. Bl\"ochl, Phys. Rev. B \textbf{50}, 17953 (1994).

\bibitem{nalkanes}
E. Londero, E.K. Karlson, M. Landahl, D. Ostrovskii, J.D. Rydberg, and E. Schr\"oder,
\textit{Desorption of n-alkanes from graphene: a van der Waals density functional
 study}, arXiv:1205.1295.

\bibitem{PAHgg}
S.D. Chakarova-K\"ack, A. Vojvodic, J. Kleis, P. Hyldgaard, and E. Schr\"oder,
New J. Phys. \textbf{12},  013017 (2010).

\bibitem{PAHgraphite}
S.D. Chakarova-K\"ack, E. Schr\"oder, B.I. Lundqvist, and
D.C. Langreth, Phys. Rev. Lett. \textbf{96}, 146107 (2006).

\bibitem{adenine}
K. Berland, S.D. Chakarova-K\"ack, V.R. Cooper, D.C. Langreth, and E. Schr\"oder,
J. Phys.: Condensed Matters \textbf{23}, 135001 (2011).

\bibitem{slidingrings}
K. Berland, T.L. Einstein, and P. Hyldgaard,
Phys. Rev. B \textbf{80}, 155431 (2009).

\bibitem{fire} E. Bitzek, P. Koskinen, F. G\"ahler, M. Moseler, and
P. Gumbsch,
Phys. Rev. Lett. \textbf{97}, 170201 (2006).

\bibitem{Ziambaras}
E. Ziambaras, J. Kleis, E. Schr\"oder, and P. Hyldgaard,
Phys. Rev. B \textbf{76}, 155425 (2007).

\bibitem{revPBE}
Y. Zhang and W. Yang, Phys. Rev. Lett. \textbf{80}, 890 (1998).

\bibitem{IJQC}
D.C. Langreth, M. Dion, H. Rydberg, E. Schr\"oder, P. Hyldgaard, and B.I. Lundqvist,
Intern. J. of Quantum Chem. \textbf{101}, 599 (2005).

\bibitem{NTgg}
J. Kleis, E. Schr\"oder, and P. Hyldgaard,
Phys. Rev. B \textbf{77}, 205422 (2008).

\bibitem{nist}NIST Computational Chemistry Comparison and Benchmark Database,
NIST Standard Reference Database Number 101,
Release 15b, August 2011, Editor: Russell D. Johnson III,
http://cccbdb.nist.gov/

\bibitem{methanol} E. Schr\"oder, \textit{Methanol adsorption
on graphene}, unpublished. 

\bibitem{tait2006}
S.L. Tait, Z. Dohn\'alek, C.T. Campbell, and B.D. Kay,
J. Chem. Phys. \textbf{125}, 234308 (2006).

\bibitem{ringer} 
A. Ringer, M. Figgs, M. Sinnokrot, and C.D. Sherrill, 
J. Phys. Chem. A \textbf{110}, 10822 (2006). 

\bibitem{harris}
J. Harris, Phys. Rev. B \textbf{31}, 1770 (1985).

\bibitem{murray}
\'E.D. Murray, K. Lee, and D.C. Langreth, J. Chem.
Theor. Comput. \textbf{5}, 2754 (2009).

\bibitem{h2cu}
K. Lee, A.K. Kelkkanen, K. Berland, S. Andersson, D.C. Langreth,
E. Schr\"oder, B.I. Lundqvist, and P. Hyldgaard,
Phys. Rev. B \textbf{84}, 193408 (2011).

\bibitem{h2cu2}
K. Lee, K. Berland, M. Yoon, S. Andersson, 
E. Schr\"oder, P. Hyldgaard, and B.I. Lundqvist,
\textit{Benchmarking van der Waals Density Functionals with Experimental 
Data: Potential Energy Curves for H$_2$ Molecules on Cu(111), (100), and 
(110) Surfaces},
preprint (2012).

\bibitem{voloshina}
E. Voloshina, D. Usvyat, M. Sch\"utz, Y. Dedkov, and B. Paulus, 
Phys. Chem. Chem. Phys.  \textbf{13}, 12041 (2011).

\bibitem{li2012}
X. Li, J. Feng, E. Wang, S. Meng, J. Klime\v{s}, and A. Michaelides, 
Phys. Rev. B \textbf{85}, 085425 (2012).

\bibitem{klimes2011}
J. Klime\v{s}, D.R. Bowler, and A. Michaelides, 
Phys. Rev. B \textbf{83}, 195131 (2011). 

\end{thebibliography}
\end{document}